\documentclass[twocolumn,9pt]{article} % here we use the article class, rather than elsarticle

\usepackage[square,numbers,sort&compress,comma]{natbib}

\usepackage{amsmath}
\usepackage{amssymb}
\usepackage{caption}
\usepackage{graphicx}
\usepackage{latexsym}
\usepackage{times}
\usepackage[pagewise]{lineno}
\usepackage{hyperref}

\usepackage{tikz}
\definecolor{mplblue}{HTML}{1F77B4}
\definecolor{mplorange}{HTML}{FF7F0E}
\DeclareRobustCommand{\solidline}[1]{%
  \tikz[baseline=-0.6ex]\draw[#1, line width=1.0pt] (0,0) -- (0.6em,0);%
}

\usepackage{eso-pic} % Required to place text on the page
\usepackage{lipsum}  % For placeholder text

\newcommand{\preprintfooter}{
  \AddToShipoutPictureBG*{%  * means only on this page
    \AtTextLowerLeft{%       % Start at bottom left of text area
      \makebox[\textwidth]{% % Box to span width
        \hspace{0pt}%        % Horizontal positioning
        \raisebox{-1cm}[0pt][0pt]{% % Lower below page text
          \small\textit{Preprint submitted for review} % Your text
        }%
        \hfill%              % Pushes content right if needed
      }%
    }%
  }%
}

\topmargin - 12pt % might need to be set to 0pt for some installations
\renewenvironment{abstract}%
              {% - begin definition
               \small% - select font
               {\bfseries \abstractname}% - select font
               \par% - end a paragraph (skip \parsep)
               \vspace{10pt}% - add vertical space
              }% - complete definition

\renewcommand\abstractname{Abstract}

\newcommand{\nomenclature}% - name of command
              [1]% - number of arguments
              {% - begin definition
               \bgroup% - begin a local group
               \flushleft% - turn on flushleft option
               \small\bf% - select font
               #1% - insert title text
               \par% - end a paragraph (skip \parsep)
               \egroup% - terminate local group
              }% - complete definition

\renewcommand{\section}% - name of command
              [1]% - number of arguments
              {% - begin definition
               \bgroup% - begin a local group
               \flushleft% - turn on flushleft option
               \small\bf% - select font
               \refstepcounter{section}% - increment counter
               \arabic{section}. #1% - insert title text
               \par% - end a paragraph (skip \parsep)
               \egroup% - terminate local group
              }% - complete definition

\renewcommand{\subsection}% - name of command
              [1]% - number of arguments
              {% - begin definition
               \bgroup% - begin a local group
               \flushleft% - turn on flushleft option
               \small\em% - select font
               \refstepcounter{subsection}% - increment counter
               \arabic{section}.% - insert title text
               \arabic{subsection}. #1% - insert title text
               \par% - end a paragraph (skip \parsep)
               \egroup% - terminate local group
              }% - complete definition

\renewcommand{\subsubsection}% - name of command
              [1]% - number of arguments
              {% - begin definition
               \bgroup% - begin a local group
               \flushleft% - turn on flushleft option
               \small\em% - select font
               \refstepcounter{subsubsection}% - increment counter
               \arabic{section}.% - insert title text
               \arabic{subsection}.% - insert title text
               \arabic{subsubsection}. #1% - insert title text
               \par% - end a paragraph (skip \parsep)
               \egroup% - terminate local group
              }% - complete definition

  \newcommand{\acknowledgement}% - name of command
              [1]% - number of arguments
              {% - begin definition
               \bgroup% - begin a local group
               \flushleft% - turn on flushleft option
               \small\bf% - select font
               #1% - insert title text
               \par% - end a paragraph (skip \parsep)
               \egroup% - terminate local group
              }% - complete definition

  \newcommand{\sectionbib}% - name of command
              [1]% - number of arguments
              {% - begin definition
               \bgroup% - begin a local group
               \flushleft% - turn on flushleft option
               \small\bf% - select font
               #1% - insert title text
               \par% - end a paragraph (skip \parsep)
               \egroup% - terminate local group
              }% - complete definition

\begin{document}
\preprintfooter % Call the footer on the first page
% -------------------------------------------------------------------- %
% -------------------------------------------------------------------- %
% -------------------------------------------------------------------- %

% -------------------------------------------------------------------- %

\small
\baselineskip 10pt

% -------------------------------------------------------------------- %
% -------------------------------------------------------------------- %
% -------------------------------------------------------------------- %
\setcounter{page}{1}
% -------------------------------------------------------------------- %
\title{\LARGE \bf A mapping-based projection of detailed \\ 
                kinetics uncertainty onto reduced manifolds}

\author{{\large Vansh Sharma$^{a,*}$, Shuzhi Zhang$^{a}$, Rahul Jain$^{a}$, Venkat Raman$^{a}$}\\[10pt]
        {\footnotesize \em $^a$Department of Aerospace Engineering, University of Michigan, Ann Arbor, MI 48109-2102, USA}\\[-5pt]}
        % {\footnotesize \em $^b$Author affiliation 2}\\[-5pt]
        % {\footnotesize \em Continue the list of affiliations as needed, with one per line}}

\date{}  %%% Leave as is, do not add date;

% -------------------------------------------------------------------- %
% -------------------------------------------------------------------- %
% -------------------------------------------------------------------- %
\twocolumn[\begin{@twocolumnfalse}
\maketitle
\rule{\textwidth}{0.5pt}
\vspace{-5pt}

\begin{abstract} % 100 to 300 words.
Propagating uncertainties introduced by chemical reaction rate parameters to high-fidelity numerical simulations of complex combustion devices is necessary to ascertain impact on computational predictions. However, the high cost of detailed computations combined with the need to conduct multiple simulations to propagate uncertainty makes such an estimation computationally challenging. In order to reduce the computational cost, a two-step framework for quantifying uncertainty introduced by detailed chemical kinetics model parameters using reduced chemistry models is developed here. First, reduced-manifold states are uniquely reconstructed in full-composition space by following trajectories at an unburnt mixing state and integrating forward to a prescribed progress variable constraint. Second, parametric uncertainty is propagated by sampling perturbed rate coefficients from mechanism covariance matrices and integrating each realization to the target state, yielding uncertainty maps for reduced-space quantities. 
The method is applied in two configurations: a subsonic multi-tube combustor with interacting jet flames and recirculation, and a three-dimensional reacting high-speed flowpath. Uncertainty-instrumented estimated are reported for a trajectory time (time for the reconstructed unreacted mixture to reach the local target state) and for the time to equilibrium, revealing order-of-magnitude spatial variations driven by mixing, stratification, and residence-time effects. The largest relative variability occurs in low-to-intermediate temperature regimes associated with induction and the onset of heat release, where branching-related chemistry amplifies sensitivity, particularly away from stoichiometric conditions.
The method provides a scalable route to spatially resolved, physically interpretable chemistry-UQ for practical reacting-flow simulations.
\end{abstract}

\vspace{10pt}

\parbox{1.0\textwidth}{\footnotesize {\em Keywords:} Conditional stochastic projection; Uncertainty quantification; Uncertainty propagation; Chemical Kinetics; Low-dimensional manifold;}
\rule{\textwidth}{0.5pt}
*Corresponding author.
\vspace{5pt}
\end{@twocolumnfalse}] 

% \linenumbers
\section{Introduction\label{sec:introduction}} \addvspace{10pt}

Quantifying uncertainties from chemistry models is necessary to establishing the credibility of reacting flow simulations \cite{raman2019emerging, mueller2013chemical, ji2018shared, wang2021active, cai2024exploiting}. While techniques for estimating the uncertainty of reaction rate parameters has been explored widely \cite{wang2015combustion,braman2013bayesian, frenklach_uq}, the use of such estimates in forward propagation through a reacting flow solver has not been explored widely due to the inherent computational expense \cite{najm2009uncertainty,owen2017comparison, ji2019quantifying, dannert2022investigations}. Studies have explored the development of uncertainty quantification (UQ) techniques integrated with dimensionality-reduction frameworks~\cite{vajda1985principal, pepiot2008efficient, ckLaw2014skeletal,braman2013bayesian}, while dedicated response surface methods for kinetic uncertainty analysis have also been proposed in recent years~\cite{najm2009uncertainty, tomlin2014evaluation, sheen2011method, su2021uncertainty}.

A key challenge occurs when reduced chemistry models are used in reacting flow calculations to accelerate the computations. Since uncertainty estimates are often formulated for single reactions or for detailed chemistry models, a reliable estimate of rate uncertainties is often missing for such simplified chemistry models. Since mechanism reduction does not exclusively eliminate pathways and hence reactions, but may also introduce new ``lumped`` reactions or steady-state assumptions, there does not exist a one-to-one mapping between full-mechanism uncertainty estimates and those of the reduced mechanism. In this regard, Oh et al.~\cite{oh2023fast} propose a hybrid response-surface network for inverse UQ using a neural network that constrains (global parameter optimization) kinetic rate uncertainties in complex mechanisms with ignition delay time (IDT) data. Su et al.~\cite{su2021uncertainty} quantify the uncertainty of the kinetic parameters during the reduction of the mechanism by combining the sensitivity and the active subspace analyzes with a PCE surrogate and introduce a `transition state' to separate the changes due to the truncation of the parameters from the reaction coupling. Notably, these prior studies typically assume independent rate-parameter uncertainties (i.e., a diagonal covariance with no cross-correlations), which limits transferability of the uncertainty model across coupled sub-mechanisms within the detailed mechanism.

Relative to the previously discussed kinetic UQ approaches, the present work targets a different—and largely unmet—need: \emph{constructing uncertainty metrics around reduced-space operating points without requiring multiple CFD runs}.
The work in \cite{oh2023fast} focuses on inverse UQ, but does not directly yield local uncertainty fields around the operating states produced by a reduced-mechanism simulation. Similarly, \cite{su2021uncertainty} quantifies uncertainty changes during mechanism reduction, but does not produce CFD-ready uncertainty maps. In contrast, the proposed framework conditions on reduced-space CFD states, reconstructs them uniquely to the detailed mechanism, and propagates a correlated kinetic prior via the detailed-mechanism covariance to generate spatially resolved variability estimates for ignition-relevant quantities without requiring prohibitively expensive full-field ensemble CFD. This provides a practical bridge between reduced-chemistry simulations and detailed-mechanism uncertainty, while also highlighting regimes where reduction-induced discrepancies warrant further scrutiny.

% {\color{red} the next para does not lead from this. What is wrong with the methods used here? What are we proposing that is different?}

% {\color{red} This should not be in the submitted file; lead from previous statement to say what has been done}
% \emph{What does this study aim to add beyond reduced-mechanism fidelity and efficiency?} It is clear from the above discussion that chemical kinetic mechanisms are typically developed and validated to reproduce targeted datasets, yet the resulting models are often deployed deterministically—without routinely carrying forward quantified uncertainty in their rate parameters. This gap persists and can widen during mechanism reduction: reduced mechanisms are commonly optimized for fidelity and cost, but uncertainty information is rarely propagated or reconstructed alongside the reduction. As a result, simulations may appear predictive while lacking a measure of chemistry-driven confidence, particularly at operating conditions beyond the calibration set. The availability of a detailed-mechanism covariance matrix enables this missing layer of rigor: the procedure in this work utilizes these correlated parameter uncertainties to construct uncertainty metrics around reduced-space operating points without requiring prohibitively expensive full-field ensemble CFD.

\begin{figure}[!ht]
    \centering
    \includegraphics[width=0.99\linewidth]{ 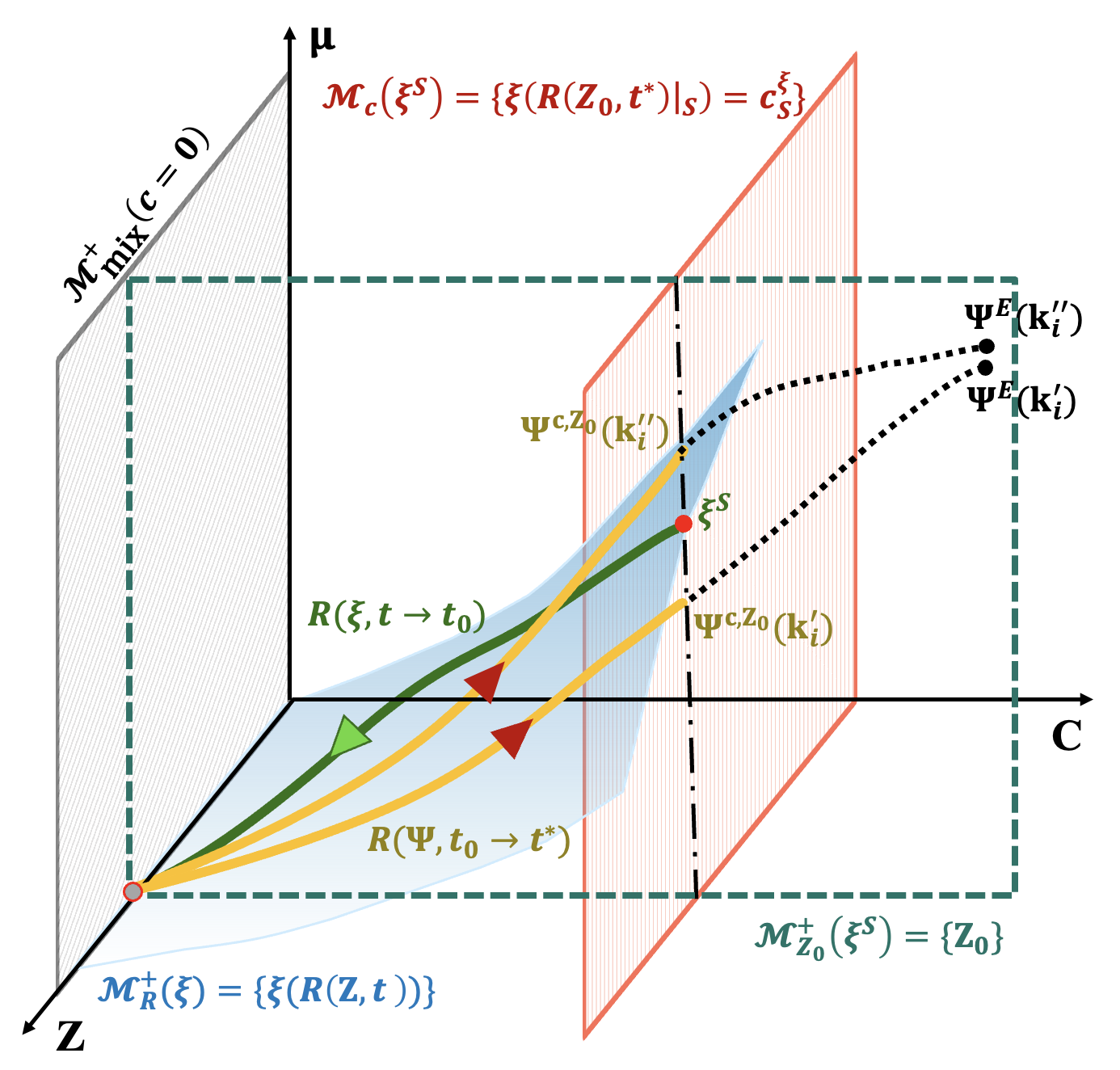}
    \caption{\footnotesize Conceptual sketch of the proposed trajectory mapping in composition space. The composition space is spanned by $Z$, $c$, and an unconstrained species $\mu$ (not frozen during reaction evolution). The blue-shaded surface $\mathcal{M}_R^{+}(\xi)$: Reduced-mechanism reaction manifold. 
    Red-shaded plane $\mathcal{M}_c(\xi^S)$: iso-$c$ manifold at state $S$ for $\xi$.
    Green trajectory $R(\xi, t\rightarrow t_0)$: reconstruction from a given state $S$ of $\xi$.
    Yellow trajectories $R(\Psi, t_0\rightarrow t^*)$: forward integration of $\Psi$ from $Z_0$ with perturbed reaction rate coefficients $k_i$ towards $\mathcal{M}_c(\Psi^E)$, and reach the target composition $\phi^{\Psi^c}(k_i)$. Dashed black lines: relaxation of different reaction trajectories toward equilibrium ($E$).}
\label{fig:mixfrac_traj}
\end{figure}

Our approach is based on a mapping between the reduced and detailed thermochemical space. To quantify uncertainty in the reduced space while using rate uncertainty from the detailed mechanism, a two-step procedure is used here: (i) reconstruction—map reduced-manifold states into the full composition space (\S\ref{sec:reconstruction}); (ii) sampling—integrate detailed reaction trajectories with perturbed rate coefficients drawn from the detailed mechanism’s covariance matrices (\S\ref{sec:sample}). This propagates parametric uncertainties and evaluates their impact on key trajectory targets in the reduced space. \emph{} Reconstruction follows the attracting-manifold assumption (cf. ICE–PIC~\cite{ren2005species}); however, unlike ICE–PIC, here the trajectories are anchored at an unburnt mixing state defined by a conserved mixture fraction and are integrated forward along the reaction mapping until the progress variable reaches the target state, ensuring a unique mapping. This eliminates intermediate search steps, improving efficiency while preserving elemental conservation and manifold invariance (see Fig.~\ref{fig:mixfrac_traj}). The following section describes the method ($\S$\ref{sec:method}) and, later, we present two use cases demonstrating the proposed approach ($\S$\ref{sec:results}).

% \begin{figure}[!ht]
%     \centering
%     \includegraphics[width=0.85\linewidth]{ Diagram.png}
%     \caption{\footnotesize Conceptual sketch of the proposed trajectory mapping in composition space. The composition space is spanned by $Z$, $c$, and an unconstrained species $\mu$ (not frozen during reaction evolution). The blue-shaded surface $\mathcal{M}_R^{+}(Z)$: forward-reacted manifold. 
%     Red-shaded plane $\mathcal{M}_c(\xi^e)$: iso-$c$ manifold.
%     % , where all states share the same progress variable value as the evolved state $R(\Psi_0, t^*)$. 
%     The green trajectory $R(\xi, -t)$ indicates the \emph{backward integration} from a given equilibrium (or target) state $\xi$.
%     % , tracing back along the reaction path to reconstruct its unburnt composition. 
%     The orange trajectories $R(\Psi, t^*)$ represent \emph{forward integrations} from initial unburnt state with perturbed reaction rate coefficients $k_i$ towards $\mathcal{M}_c(\Psi^e)$, and reach the target composition $\phi^\psi(k_i)$. The dashed black lines depict the relaxation of different reaction trajectories toward equilibrium.}
% \label{fig:mixfrac_traj}
% \end{figure}

% -------------------------------------------------------------------- %

\section{Methodology\label{sec:method}} \addvspace{10pt}

% The initial state of the mixture, computed using the reduced mechanism, is assumed to be a state along the equilibrium state trajectory. To estimate uncertainty bounds on the reduced mechanism using the detailed mechanism, a two-step approach is developed. First, the reduced manifold states are reconstructed in the full composition space as detailed in \ref{sec:reconstruction}. Second, detailed reaction trajectories are computed using perturbed reaction rate coefficients drawn from the specified covariance matrices of the detailed mechanism (see \ref{sec:sample}). This two-step process enables the propagation of parametric uncertainties and the evaluation of their impact on key trajectory targets in the reduced space.

\subsection{Species reconstruction\label{sec:reconstruction}}\addvspace{10pt}
This study assumes a gas-phase mixture ($\phi_g$) of $n_s$ chemical
species composed of $n_e$ elements ($\mathbf{E}$ is elemental matrix) in a closed, homogeneous and ideal gas reacting system. The thermochemical state of the mixture (at any time) is completely characterized by the pressure $p$, the temperature $T$, and the $n_s$-vector $\chi$ of the specific moles of the species, $\phi_g = (p,T,\chi)$. Say $p$ and $T$ are held constant, the state is given by $\chi$. Due to chemical reactions, the composition $\chi(t)$ evolves in time $t$ according to the set of ordinary differential equations (ODEs) as \(\tfrac{d}{dt}\mathbf{\chi}(t) = \mathbf{S}_\Psi(\mathbf{\chi}(t))\) \text{(1)}, where $\mathbf{S}_\Psi\in\mathbb{R}^{n_s}$ is the vector of chemical production rates of the detailed mechanism ($\Psi$). A similar equation can be written using the production rate vector of the reduced mechanism ($\xi$), $\mathbf{S}_\xi\in\mathbb{R}^{n_r}$. 
The production rate vector $\mathbf{S}$ satisfies: (1) \textit{Elemental conservation}: $\mathbf{E}^\top \mathbf{S} = \mathbf{0}$, ensuring that atomic elements are conserved throughout the reaction; (2) \textit{Reactivity}: The mixture remains reactive except at chemical equilibrium, i.e., $\|\mathbf{S}\| > 0$ for all $\boldsymbol{\chi} \neq \boldsymbol{\chi}^{eq}$; (3) \textit{Thermodynamic consistency}: The Gibbs free energy $G(\boldsymbol{\chi})$ is a strictly decreasing function along any reaction trajectory, $S^T\frac{dG}{dt} < 0$ for $\boldsymbol{\chi} \neq \boldsymbol{\chi}^{eq}$, with $\chi \to \chi^{eq}$ as $t \to \infty$; and (4) \textit{Realizability}: Consistent with the law of mass action, evolution of species concentrations is restricted to the realizable region of the composition space, i.e., $\chi_i(t) \ge 0, \ \forall i, t$. These properties have been discussed in detail in~\cite{ren2006invariant}. 

The two mechanisms ($\boldsymbol{\xi}$ and $\boldsymbol{\Psi}$) are subject to the following assumptions: $\boldsymbol{\xi} \subset \boldsymbol{\Psi}$, $\mathbf{S}_\xi = \mathbf{P}\,\mathbf{S}_\Psi$ and $\boldsymbol{\xi}, \boldsymbol{\Psi}$ contain the same set of elemental types, where $\mathbf{P}$ is a projection ensuring that individual reaction rates remain consistent (values may vary) in both mechanisms. The term $\mathbf{R}(\chi,t)$ represents the reaction mapping, which is the solution to equation (1) at time $t$, starting from the initial condition $\chi$. 
Given an unburned state $\phi_u^{\xi} = (p,T, \chi
^\xi_u)$, mixture fraction $Z^{\xi}$ can be calculated as outlined in \cite{bilger1990reduced}. This state is then advanced along the $\boldsymbol{\xi}$ trajectory, and at time $\tau^\xi$, a progress variable ($c_g^{\xi}$) is defined as a linear combination of selected species compositions. Initial state for $\boldsymbol{\Psi}$ can be defined as $\phi_u^{\Psi} =  (p,T, \chi^\Psi_u(Z^{\xi}))$ where $\chi^\Psi_u(Z^{\xi}) = B\chi^\xi_u(Z^{\xi})$ with $B \in\mathbb{R}^{n_s \times n_r}$ as the constant species map to reconstruct the detailed species composition from $\chi^{(\xi)}(Z)\xrightarrow{\ B\ } \chi^{(\Psi)}(Z)$.
The system $\boldsymbol{\Psi}$ is initialized on the corresponding unburnt mixing edge \(\boldsymbol{\Psi}_0 = \boldsymbol{\Psi}_u(p,T\chi^\Psi_u)\) and advanced along the detailed trajectory \(t \mapsto \mathbf{R}(\boldsymbol{\Psi}_0, t)\) 
until the progress variable reaches the prescribed value \(c_g^{\xi}\), i.e., $c_g^{\Psi}(\mathbf{R}(\boldsymbol{\Psi}_0, t^\star)) = c_g^{\xi}$ (see Fig.~\ref{fig:mixfrac_traj}). The resulting detailed composition \(\boldsymbol{\Psi}^\star = \mathbf{R}(\boldsymbol{\Psi}_0, \tau^\star)\) satisfies both mixture-fraction and progress-variable consistency, \(Z(\boldsymbol{\Psi}^\star)=Z^{\xi}\) and \(c(\boldsymbol{\Psi}^\star)=c_g^{\xi}\), 
and thus represents the reconstructed detailed state corresponding to the reduced coordinate \(\boldsymbol{\phi}_g^{\xi}\). If \(\boldsymbol{c}\) evolves monotonically along the trajectory—i.e., if \(\mathbf{S}_\psi(\mathbf{R}(\boldsymbol{\Psi}_0, t))\) maintains a fixed sign—the root \(\tau^\star\) is unique \cite{ren2005species}.

\subsection{Uncertainty quantification\label{sec:sample}} \addvspace{10pt}
Uncertainty in reaction rates for $\boldsymbol{\Psi}$ is represented by sampling correlated Arrhenius parameters from a multivariate Gaussian distribution with covariance $\Sigma_\Psi$.  For each reaction $i$, the normalized variables $x_{A,i}=\ln(A_i/A_{0,i})/\ln f_i$ and $x_{B,i}=(B_i-B_{0,i})/(T_{\mathrm{ref}}\ln f_i)$ describe perturbations in the pre-exponential factor and activation energy.  The corresponding logarithmic rate variation is $x_i(T)=x_{A,i}-(T_{\mathrm{ref}}/T)x_{B,i}$, yielding $k_i(T)=k_{0,i}(T)\,f_i^{\,x_i(T)}$, with $\boldsymbol{x}_i=[x_{A,i},x_{B,i}]^\top$. The joint parameter vector $\mathbf{X}=[\boldsymbol{x}_1^\top,\ldots,\boldsymbol{x}_I^\top]^\top$ follows $\mathbf{X}\sim\mathcal{N}(\boldsymbol{\mu_m},\boldsymbol{\Sigma_\Psi})$, where $\boldsymbol{\Sigma_\Psi}$ encodes correlations between $A_i$ and $B_i$.  Samples are drawn by $\mathbf{X}^{(s)}=\boldsymbol{\mu_m}+\mathbf{L}\mathbf{z}^{(s)}$ with $\mathbf{L}\mathbf{L}^\top=\boldsymbol{\Sigma_\Psi}$ and $\mathbf{z}^{(s)}\sim\mathcal{N}(0,\mathbf{I})$.  
% Because $\ln k_i(T)=\ln k_{0,i}(T)+(\ln f_i)\boldsymbol{\alpha}^\top\boldsymbol{x}_i$ is linear in normal variables, $k_i(T)$ is lognormal with mean $\mu_{\ln k,i}(T)=\ln k_{0,i}(T)+(\ln f_i)\boldsymbol{\alpha}^\top\boldsymbol{\mu}_i$ and variance $\sigma_{\ln k,i}^2(T)=(\ln f_i)^2\boldsymbol{\alpha}^\top\boldsymbol{\Sigma}_i\boldsymbol{\alpha}$.  
Repeated sampling and integration of the kinetic model generate an ensemble of trajectories whose dispersion defines an uncertainty band (see Fig.~\ref{fig:mixfrac_traj}). The covariance matrix must be symmetric and positive definite. %; When reactions are independent, it is block-diagonal with per-reaction prior, which is strictly positive definite for $T_{\min}\neq T_{\max}$, ensuring consistent stochastic sampling.

% -------------------------------------------------------------------- %
\section{Results and Discussion\label{sec:results}} \addvspace{10pt}
In this section, the proposed two-step procedure is further evaluated across two distinct flow configurations. First, a subsonic configuration with partially premixed combustion and reduced geometric influence on chemical timescales is investigated to demonstrate the approach. Second, a supersonic reacting-flow case is examined to assess the impact of disparate time scales. 

\subsection{Reacting multi-tube combustor\label{sec:sim_mtm}} \addvspace{10pt}

\begin{figure}[!ht]
    \centering
    \includegraphics[width=0.99\linewidth]{ 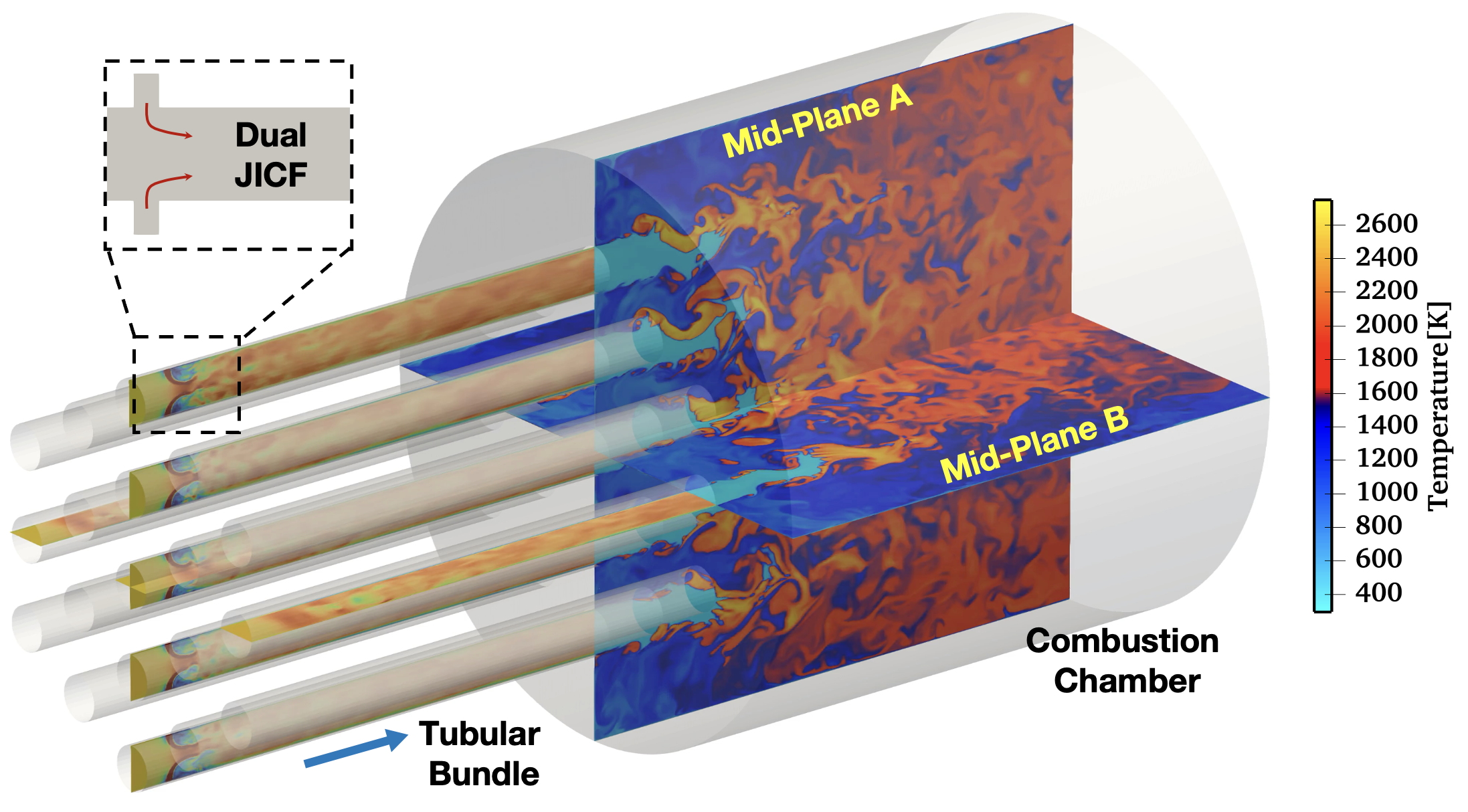}
    \caption{\footnotesize Instantaneous snapshot on the combustor mid-plane showing temperature. Blue arrows indicate the flow direction of the oxidizer stream. Red arrows depict fuel stream issuing perpendicular into the oxidizer stream (inset).}
\label{fig:MTM_setup}
\end{figure}

The numerical simulation models a multi-tube combustor with a subsonic reacting flow, where a flame, potentially under a partially-premixed combustion mode, is anchored at the exit region of a tubular bundle \cite{shuzhi_mtm}. This geometry has been validated against experimental data, and details of the computational domain are provided in \cite{shuzhi_mtm}. Two primary wall injectors, angled at 90\textdegree~introduce hydrogen fuel within each tube element, upstream of the combustion chamber. Vitiated air at 758 K and 162 m/s enters the tubular bundle following the experimental setup. Here, the chamber region of the numerical results is extracted only along the mid-section plane B (Fig.~\ref{fig:MTM_setup}). Upstream fuel injection induces mixture stratification that, along with evolving recirculation zones, modulates the flame front. This coupling promotes periodic detachment of localized reaction pockets and drives oscillatory flame dynamics, making this region a key focus for further investigation. It is important to note that this system is a subsonic deflagration case compared to the previous case.
The simulations are performed using a fully compressible Navier-Stokes solver~\cite{numericsHMM}, employing an H$_2$-O$_2$ mechanism with 9 species and 21 reactions~\cite{mueller1999flow}. In this study, cells with $T>$ 900 K are termed chemically ``active cells'' and are considered in the analysis; however, if the cell has a fuel concentration near zero, it is discarded.

\subsubsection{Trajectories with FFCM-2 for H$_2$ fuel \label{sec:one_cell_mtm}} \addvspace{10pt}

\begin{figure}[!htb]
    \centering
    \includegraphics[width=0.95\linewidth]{ 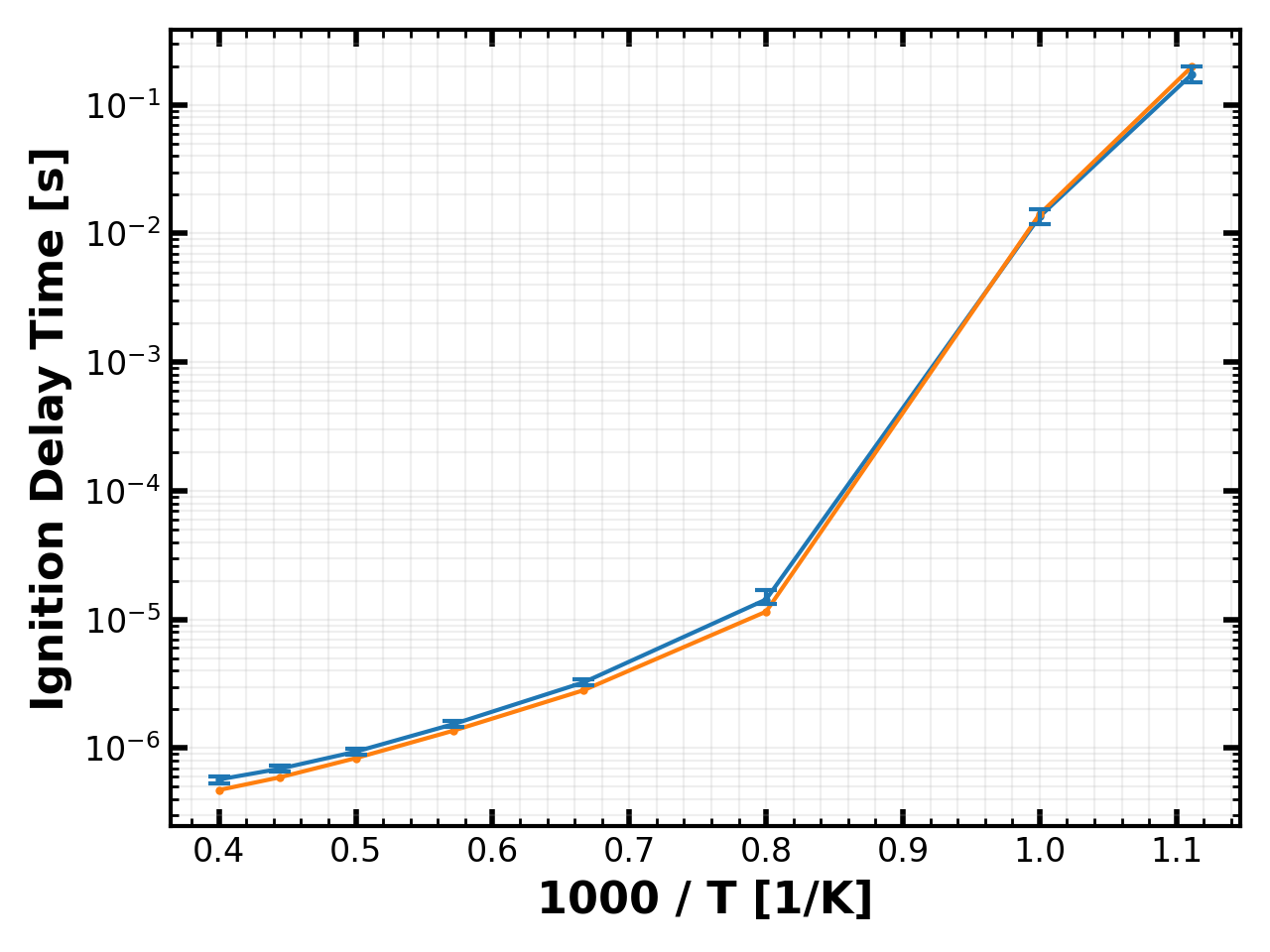}
    \caption{\footnotesize Ignition delay time (log-scale) comparison for $H_2$ at P = 10 atm and $\phi = 0.5$ as a function of inverse temperature (1000/T), Predictions from the FFCM-2 mechanism (\solidline{color=mplblue}) are compared against the H$_2$ mechanism~\cite{mueller1999flow}(\solidline{color=mplorange}); Bars indicate the FFCM-2 uncertainty band.}
\label{fig:idt_compare}
\end{figure}

Here, the FFCM-2 mechanism, along with its covariance matrix, will be used to study the H$_2$ fuel reactions. Since this is a nontrivial task, validation is first conducted by computing the ignition delay time (IDT) for FFCM-2 using only hydrogen as the fuel species.
Figure~\ref{fig:idt_compare} compares the IDT for H$_2$ fuel with FFCM-2 and H$_2$ mechanism~\cite{mueller1999flow} under the simulation conditions obtained using the $\frac{dT}{dt}|_{max}$ ignition criterion. Across the full temperature range considered, the IDT trends predicted by FFCM-2 closely match those from the H$_2$ mechanism, with differences that are small relative to the overall variation in IDT (spanning several orders of magnitude with temperature). In addition, the FFCM-2 predictions remain consistent with its reported uncertainty band (0.001--0.999 quantiles), and the alternative mechanism largely falls within this range. The strong agreement indicates that using the FFCM-2 mechanism for H$_2$ fuel reactions under these conditions is appropriate, as it reproduces comparable ignition delay behavior to the reference mechanism without introducing meaningful bias in the IDT predictions.

\begin{figure}[!htb]
    \centering
    \includegraphics[width=0.99\linewidth]{ 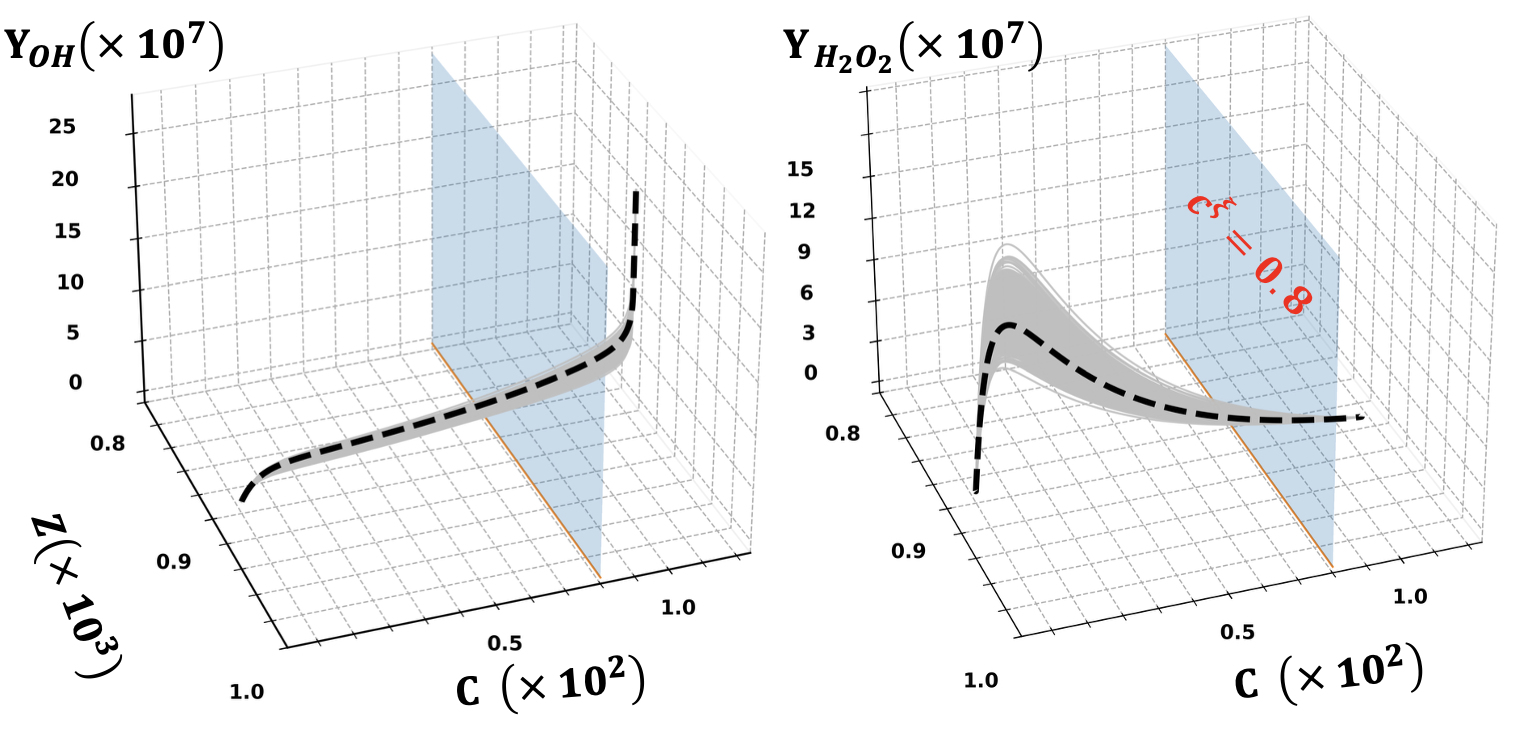}
    \caption{\footnotesize Ensemble of one-cell reaction trajectories in C, $Z$, $\mu$ space, and two different un-constrained coordinate definition. Gray curves span different sample of reaction kinetics; one typical trajectory is emphasized with a thick black dashed line. The blue plane marks  $C^{\xi}=0.8$ constraint.}
\label{fig:onecell_traj_H2}
\end{figure}

% The numerical simulations use the H$_2$-O$_2$ mechanism but for the UQ analysis, the detailed model ($\boldsymbol{\Psi}$) is the FFCM-2 model (96 species, 1054 reactions)~\cite{FFCM2}, formulated with a parameter-covariance matrix ($\boldsymbol{\Sigma_\Psi}$) that captures correlations between reaction-rate parameters.
To understand the species reconstruction method and the generation of different reaction trajectories, one cell from the numerical simulation is analyzed here. 
First, following $\S$\ref{sec:reconstruction}, an unburnt mixing state is reconstructed. Next, reaction trajectories were computed with perturbed kinetics defined by a scalar multiplier \(s = f^{x_k}\), where \(x_k\) are components of the joint parameter vector $\mathbf{X}\sim\mathcal{N}(0,\boldsymbol{\Sigma_\Psi})$~\cite{FFCM2}. Using Cantera~\cite{cantera}, perturbations uniformly rescale reaction rates for $\boldsymbol{\Psi}$: pressure-independent reactions set \(k \to s\,k\); falloff reactions use \(A_0 \to sA_0\) and \(A_\infty \to sA_\infty\), preserving the reduced pressure and the falloff shape; The PLOG type applies the same $s$ to every node’s \(A_m\). Third-body efficiencies were modified as  $[M]_{\mathrm{eff}}=\sum_s \eta_s [s]$, shifting the pressure dependence (and the low-pressure branch) without altering the high-pressure limit. The ensemble of 500 different sampled reaction trajectories for a representative cell is shown in Fig.~\ref{fig:onecell_traj_H2}. To map the results to the simulation manifold $\boldsymbol{\xi}$, the reaction process was evaluated using the progress variable $C$ = Y$_{\mathrm{OH}}$ + Y$_{\mathrm{H_2O}}$, which is defined based on the original cell state. The unconstrained coordinate (representing species that are not frozen during reaction evolution) is defined as the mass fraction of the intermediate species. The intermediate species OH begins with a low radical concentration and shows a flat increase at low reaction progress levels. As the reaction progresses, the path remains on a long, gently curved low-Y$_{\mathrm{OH}}$ branch for most of its extent. Near the high-$C$ end, the trajectory turns sharply and rises almost vertically, indicating a rapid increase in Y$_{\mathrm{OH}}$ with comparatively small changes in $C$ and $Z$. At the same time, the trajectory for ${\mathrm{H_2O_2}}$ begins with an almost vertical rise at relatively low $C$, indicating a rapid build-up of $\mathrm{H_2O_2}$ while $C$ changes only weakly; as $C$ increases into an intermediate range, the path reaches a peak (or near-plateau) with Y$_{\mathrm{H_2O_2}}\approx Y_{\max}$, after which it transitions smoothly into a long declining tail in which Y$_{\mathrm{H_2O_2}}$ decreases gradually as $C$ continues to increase. In contrast to Y$_{\mathrm{OH}}$, which tends to exhibit a delayed and abrupt jump, the evolution of Y$_{\mathrm{H_2O_2}}$ is more progressive and continuous, following a build-up $\rightarrow$ peak $\rightarrow$ decay pattern. The thermochemical states at the intersection point of each reaction trajectory with the plane \(C = c^{\xi}\) are used for the next analysis.

\subsubsection{Local trajectory time \label{sec:simRes_MTM}} 
\begin{figure}[!ht]
    \centering
    \includegraphics[width=0.99\linewidth]{ 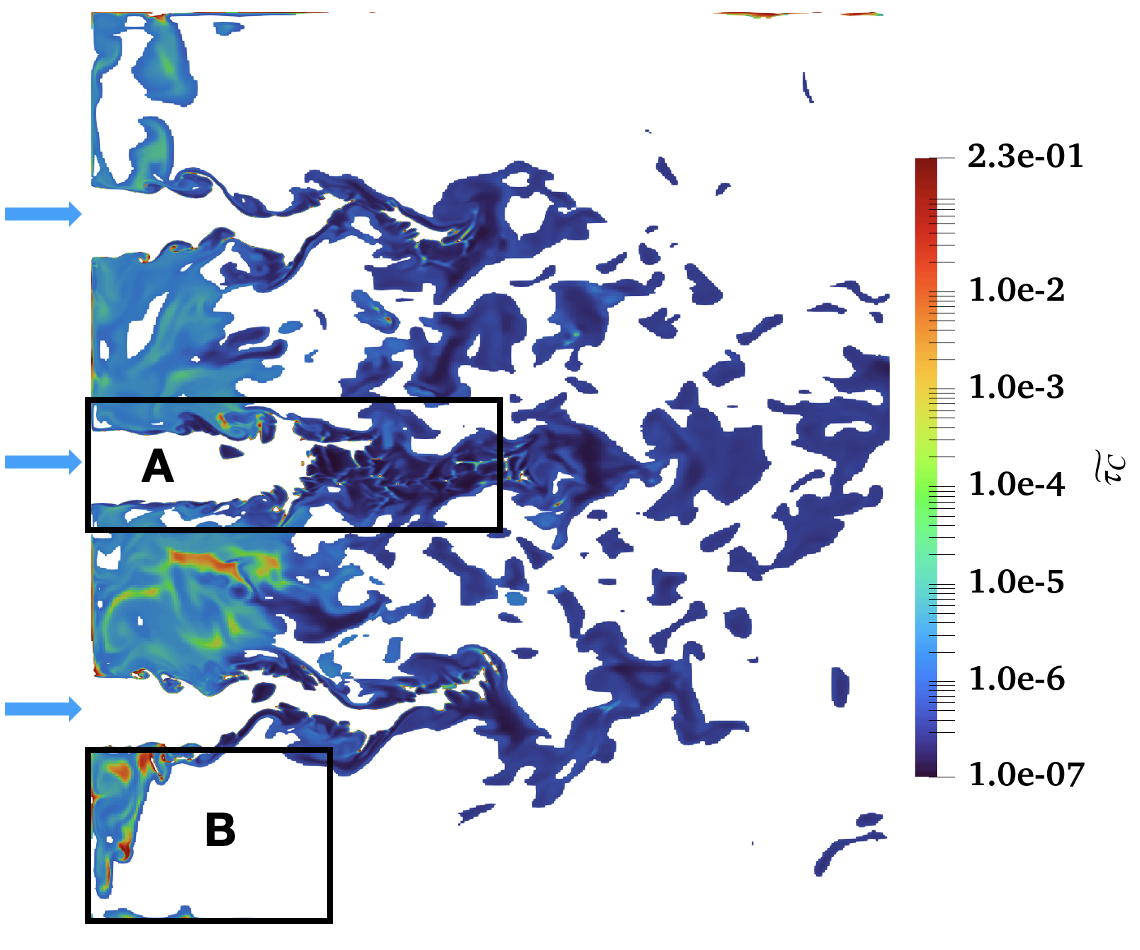}
    \caption{\footnotesize Instantaneous snapshot on the combustor mid-plane B showing time scale $\widetilde{\tau_{C}}$ in log-scale computed from $\Psi$. Blue arrows indicate the flow direction of the partially premixed stream. Region A represents the flame front, while region B represents the recirculation zone.}
\label{fig:MTM_tau_C}
\end{figure}

\addvspace{10pt}
In order to assess the UQ framework, we utilize the following performance metric: we define a time $\boldsymbol{\tau_C}$, which is the time taken by a local unreacted fuel-air mixture (based on mixture fraction) to reach the local thermodynamic state characterized by the progress variable. The initial conditions are reconstructed using $Z^{\xi}$ from each computational cell in the domain, and the integration is carried out in a constant-volume reactor for $\boldsymbol{\Psi}$, similar to the analyzes presented above for the mapping procedure. A total of 100 samples are generated for each of the 90000+ cells. Fig.~\ref{fig:MTM_tau_C} shows this mean chemical timescale ($\widetilde{\tau_C}$) on the combustor mid-plane B. Interactions between the discrete jet flames (entering the combustion chamber) and the recirculation flow generate order-of-magnitude variations in $\widetilde{\tau_C}$, reflecting ongoing mixing between mixtures of different local compositions. Along the flame front, the chemical timescale exhibits pronounced spatial variability. Regions characterized by near-stoichiometric conditions display short chemical timescales on the order of $10^{-6}\,\mathrm{s}$, indicative of rapid reaction kinetics, consistent with the behavior observed for the non-premixed diffusion flame front discussed later in \S\ref{sec:simResults}. In contrast, portions of the flame front just downstream of the tube exit are associated with substantially longer chemical timescales, reaching $\mathcal{O}(10^{-2})\,\mathrm{s}$, which can be attributed to fuel stratification within the main flow. This wide range of chemical timescales along the flame front underscores the partially premixed nature of the combustion process, wherein both kinetically fast, diffusion-controlled reactions and slower, mixing-limited reactions coexist. Further downstream, the characteristic chemical timescale $\widetilde{\tau_C}$ decreases to approximately $10^{-7}\,\mathrm{s}$, consistent with a flow region dominated by fully reacted, post-flame products and rapid radical recombination kinetics occurring under relatively low local strain rates. In contrast, within the inter-jet recirculation zone in the near vicinity of the tube exit, interactions between the jet and the recirculating flow generate strong spatial variations in strain rate and enhance local fuel stratification, leading to elevated chemical timescales and a significantly broader distribution. In this region, the chemical timescale spans several orders of magnitude, ranging from $\sim 0.2\,\mathrm{s}$ to $10^{-4}\,\mathrm{s}$, reflecting the combined effects of strain-rate-induced residence-time reduction and mixing-limited chemistry.

% \begin{figure}[!ht]
%     \centering
%     \includegraphics[width=0.99\linewidth]{ MTM_tau_C_midPlaneA.png}
%     \caption{\footnotesize Instantaneous snapshot on the combustor mid-plane A showing time scale $\widetilde{\tau_{C}}$ in log-scale computed from $\Psi$. Blue arrows indicate the flow direction of the partially premixed stream.}
% \label{fig:MTM_tau_C_planeA}
% \end{figure}

\begin{figure}[!ht]
    \centering
    \includegraphics[width=0.99\linewidth]{ 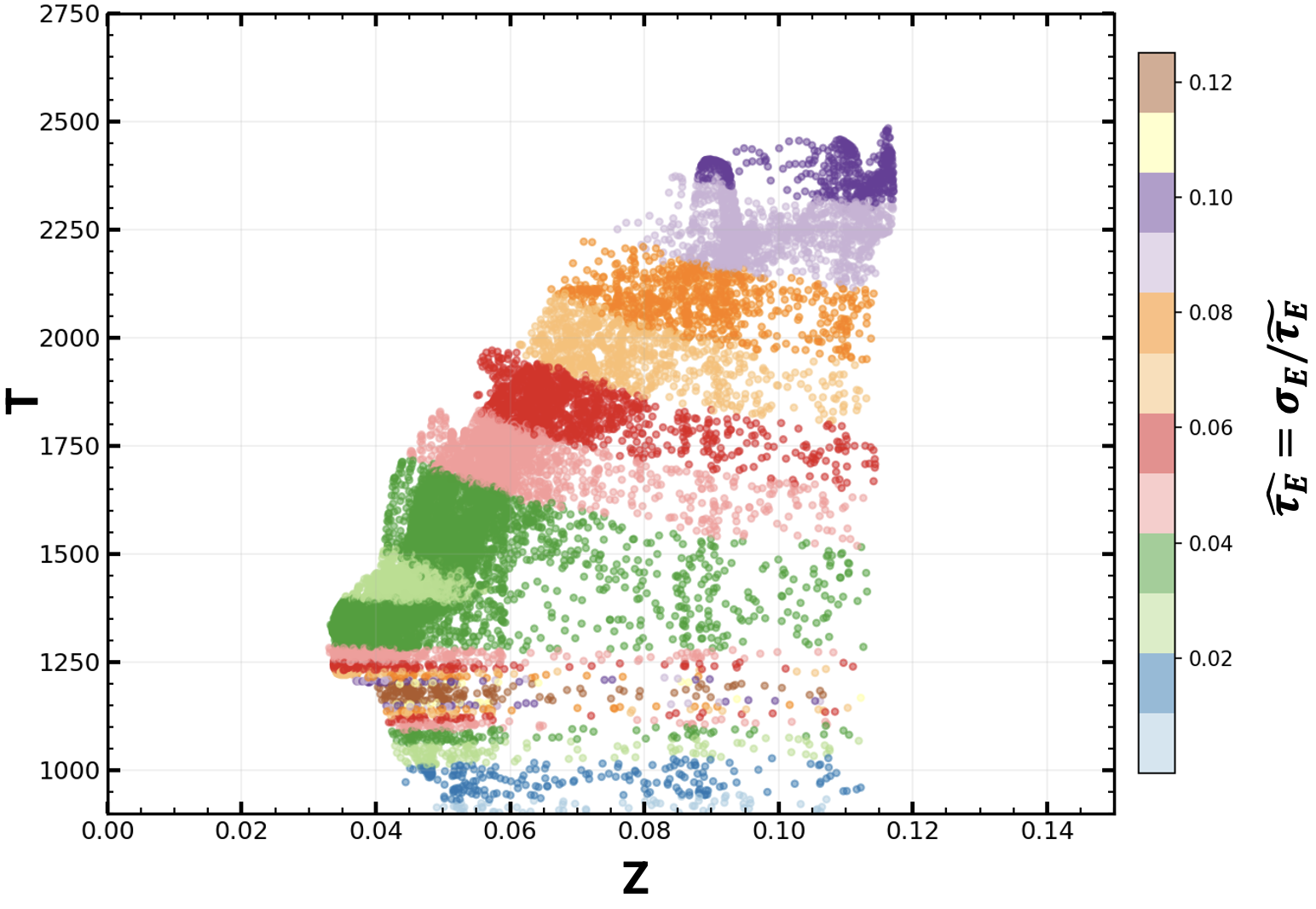}
    \caption{\footnotesize Scatter plot of T versus Z colored by $\widehat{\tau}_E$ on the flame front (Region A in Fig.~\ref{fig:MTM_tau_C}). }
\label{fig:MTM_tau_eq_center}
\end{figure}

\begin{figure}[!ht]
    \centering
    \includegraphics[width=0.99\linewidth]{ 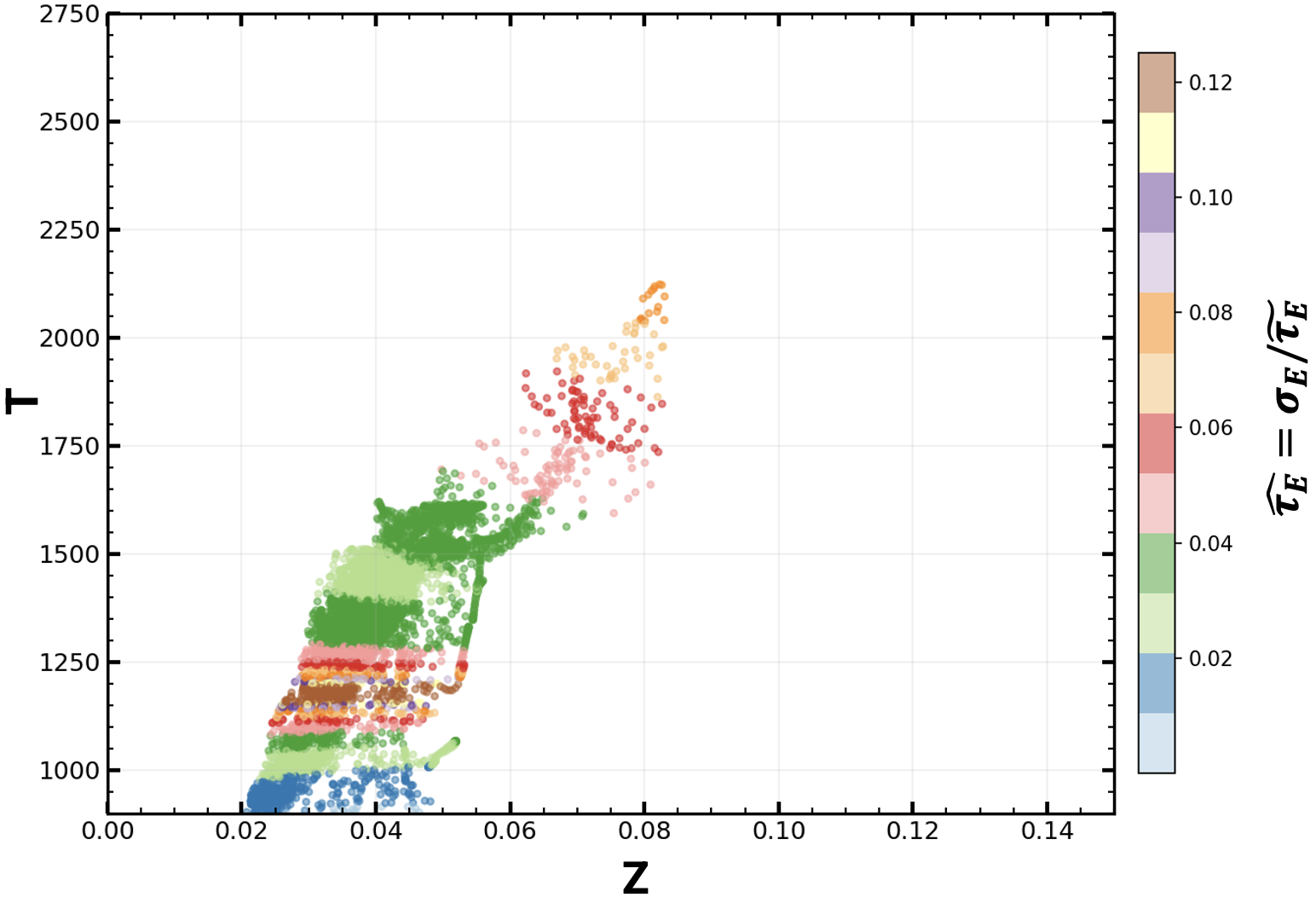}
    \caption{\footnotesize Scatter plot of T versus Z colored by $\widehat{\tau}_E$ inside recirculation zone (Region B in Fig.~\ref{fig:MTM_tau_C}).}
\label{fig:MTM_tau_eq_recirculation_zone}
\end{figure}

We extend the UQ analysis to the time-to-equilibrium. Each cell is assumed to relax via a constant-UV (internal energy and volume) equilibration process. We note that other systems may require different equilibration assumptions (e.g., UV vs. HP). For every cell, we integrate a constant-volume reactor with perturbed reaction rates and record the time at which T=T$_{UV}$, repeating this for 100 samples. Figures~\ref{fig:MTM_tau_eq_center} and \ref{fig:MTM_tau_eq_recirculation_zone} show the local temperature $T$ as a function of mixture fraction $Z$, with marker color denoting the normalized variability of the local timescale $\widehat{\tau}_E=\sigma_{\tau_E}/\widetilde{\tau}_E$. The two panels correspond to different regions—the flame front and the recirculation zone—and exhibit distinct thermochemical coverage and scatter. The flame-front samples span a broad range of mixture fractions (extending to substantially higher $Z$) and reach higher temperatures (up to $\sim 2500$ K), indicating strong heat release and intermittent exposure to richer and near-stoichiometric pockets. Correspondingly, this plot shows substantial spread in $T$ at fixed $Z$ and a wider distribution of $\widehat{\tau}_E$, reflecting stronger local stratification and more intense, unsteady reaction. By comparison, the recirculation-zone samples are confined to leaner conditions (lower $Z$) and a narrower temperature range (peaking near $\sim 2100$ K). The points cluster more tightly along a mixing/ignition trajectory, consistent with enhanced dilution by recirculated products, milder reaction, and reduced thermochemical intermittency relative to the flame front. 

Both on the flame front and in the recirculation zone, $\widehat{\tau}_E$ varies by approximately $2\%$–$18\%$, with the largest dispersion occurring in the low-to-intermediate temperature regime corresponding to induction and the onset of heat release. In this regime, we found that two chain-propagation reactions,
$\mathrm{H} + \mathrm{O_2},(+\mathrm{M}) \rightleftharpoons \mathrm{HO_2},(+\mathrm{M})$ (R1) and
$\mathrm{H_2} + \mathrm{OH} \rightleftharpoons \mathrm{H} + \mathrm{H_2O}$ (R2),
together with the chain-branching reaction
$\mathrm{H} + \mathrm{O_2} \rightleftharpoons \mathrm{O} + \mathrm{OH}$ (R3),
govern the approach to equilibrium.

Consistent with findings in~\cite{keromnes2013experimental}, under fuel-lean conditions, the chemical state is particularly sensitive to R2. Moreover, as shown in~\cite{o2004comprehensive}, the IDT exhibits strong sensitivity to R1 in the low-temperature regime near $1000~\mathrm{K}$. In addition, as discussed in~\cite{keromnes2010ignition}, the R1 pathway is strongly pressure dependent and is enhanced at elevated pressures, thus isolating $\mathrm{H}$ atoms into $\mathrm{HO_2}$ and reducing their availability for the chain-branching pathway R3, which in turn effectively modulates the overall branching propensity. In the intermediate-temperature range, the branching reaction R3 is frequently identified as one of the most sensitive elementary steps~\cite{shimizu2011updated}. Collectively, these findings suggest that the elevated variability in $\widehat{\tau}_E$ in the low-to-intermediate temperature regime is attributable to the initiation and competition of branching-related chemistry and its pressure-modulated coupling to propagation pathways.

A key inference that may be drawn from the analysis is that pronounced deviation in the low-to-intermediate temperature regime is expected to influence ignition-kernel evolution in the presence of relatively cooler and stratified reactant streams. This is important for studies focused on kernel growth or even auto-ignition behavior, as this may alter early transient flame development. During the initial phase of the simulation, when mixing and thermochemical perturbations are small but consequential, modest fluctuations in local composition and temperature can therefore yield disproportionately large changes in the effective chemical timescale. At elevated temperatures, the hydrogen chemistry considered here exhibits comparatively similar pathway evolution, even in hot states; consequently, $\widehat{\tau}_E$ remains bounded (typically $\lesssim 0.1$). 

When comparing these results along low and high $T$ regime, the variation in the chemical timescale at low temperatures is dictated by R1, where radical removal tends to control the formation/consumption of HO$_x$ and therefore uncertainties in temperature/pressure can strongly amplify the deviation in timescales~\cite{burke2012comprehensive}. However, as $T$ increases, the dominant flux shifts toward the chain-branching step $\mathrm{H+O_2\rightleftharpoons O+OH}$ (R3) as tested in Cantera~\cite{cantera}; indeed, Li et al.~\cite{li2004updated} reported that above $\sim 2000\,\mathrm{K}$, the competing association channel becomes insignificant relative to this branching reaction. However, predictions of hydrogen oxidation (e.g., ignition delay and flame propagation) can still exhibit appreciable sensitivity to the remaining high-$T$ controlling reactions and their rate parameters (most notably $\mathrm{H+O_2\rightarrow O+OH}$ and coupled HO$_x$ pathways)~\cite{burke2012comprehensive}. Therefore, the smaller variation observed at high $T$ should be interpreted primarily as an artifact of the \emph{current} uncertainty model (i.e., the assumed FFCM-2 covariance matrix $\boldsymbol{\Sigma}$ and its correlations/weights, which can decrease variance propagation when high-$T$ branching-rate parameters are treated as strongly correlated), rather than as a physical statement that hydrogen kinetics are ``not sensitive" at high temperatures.

\subsection{Reacting supersonic flow\label{sec:simulation}} 
\begin{figure}[!ht]
    \centering
    \includegraphics[width=0.95\linewidth]{ 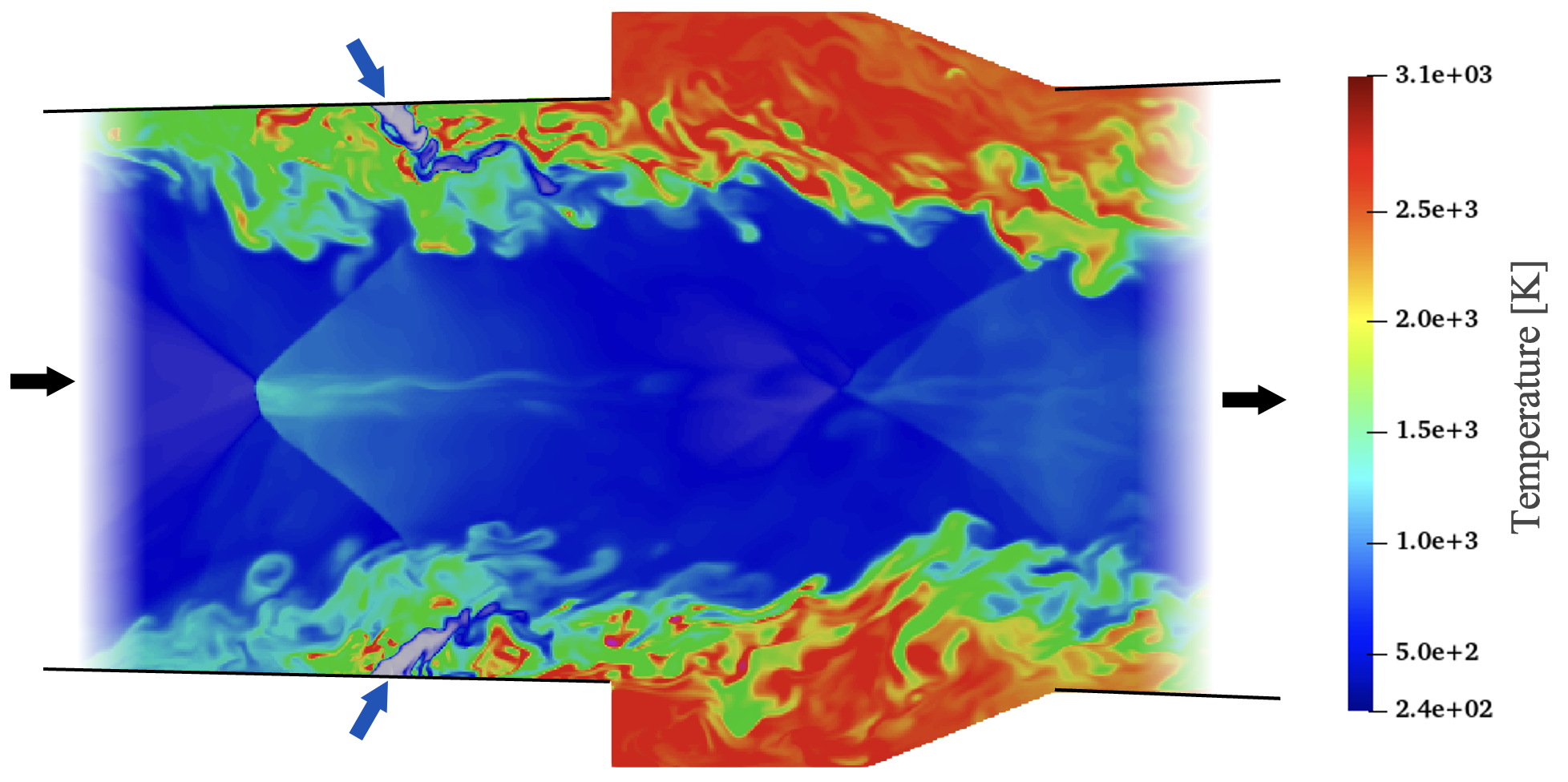}
    \caption{\footnotesize Instantaneous temperature contour along the mid-section plane of the combustor. Black arrows: direction of crossflow; Blue arrows: flow direction of fuel injection.}
\label{fig:scrjet}
\end{figure}

\addvspace{10pt}
% \subsubsection{Reacting supersonic flow}
The second applied case models a high-speed reacting flow, where a non-premixed injection based flame is anchored by a recirculation region \cite{jain2025flame}. Details of the computational domain are provided in \cite{jain2025flame,afrl_aiaapaper}. Primary wall injectors angled at 30\textdegree~introduce ethylene fuel upstream of the cavity to improve mixing and combustion. Vitiated air (with CO$_2$ and H$_2$O) at Mach 5 enters the domain, consistent with a high-enthalpy experimental inflow setup (Fig.~\ref{fig:scrjet}). The current study uses a specific region of the numerical results extracted along the mid-section plane of the combustor. This region contains oblique and normal shocks interacting with the fuel columns, while deflagration is observed within the recessed cavity. 
Simulations are performed using a fully compressible Navier-Stokes solver~\cite{numericsHMM} with the FFCMY-1 mechanism~\cite{ffcmy9reduced30_NOx_mod}. Similar to $\S$\ref{sec:sim_mtm}, cells with $T>$850K are termed chemically ``active cells'' and are considered in the analysis; however, if a cell has a fuel concentration near zero, it is discarded.

\subsubsection{Ensemble of trajectories\label{sec:singleCell}} \addvspace{10pt}

In this case, the species reconstruction method and the generation of different reaction trajectories is minimally different from $\S$\ref{sec:one_cell_mtm}.
The numerical simulations use the FFCMY-1 mechanism (30 species, 231 reactions) which is a reduced variant of the detailed FFCM-2 model ($\boldsymbol{\Psi}$). We use the same parameter-covariance matrix ($\boldsymbol{\Sigma_\Psi}$) that captures the correlations between reaction-rate parameters for FFCM-2.

\begin{figure}[!htb]
    \centering
    \includegraphics[width=0.95\linewidth]{ 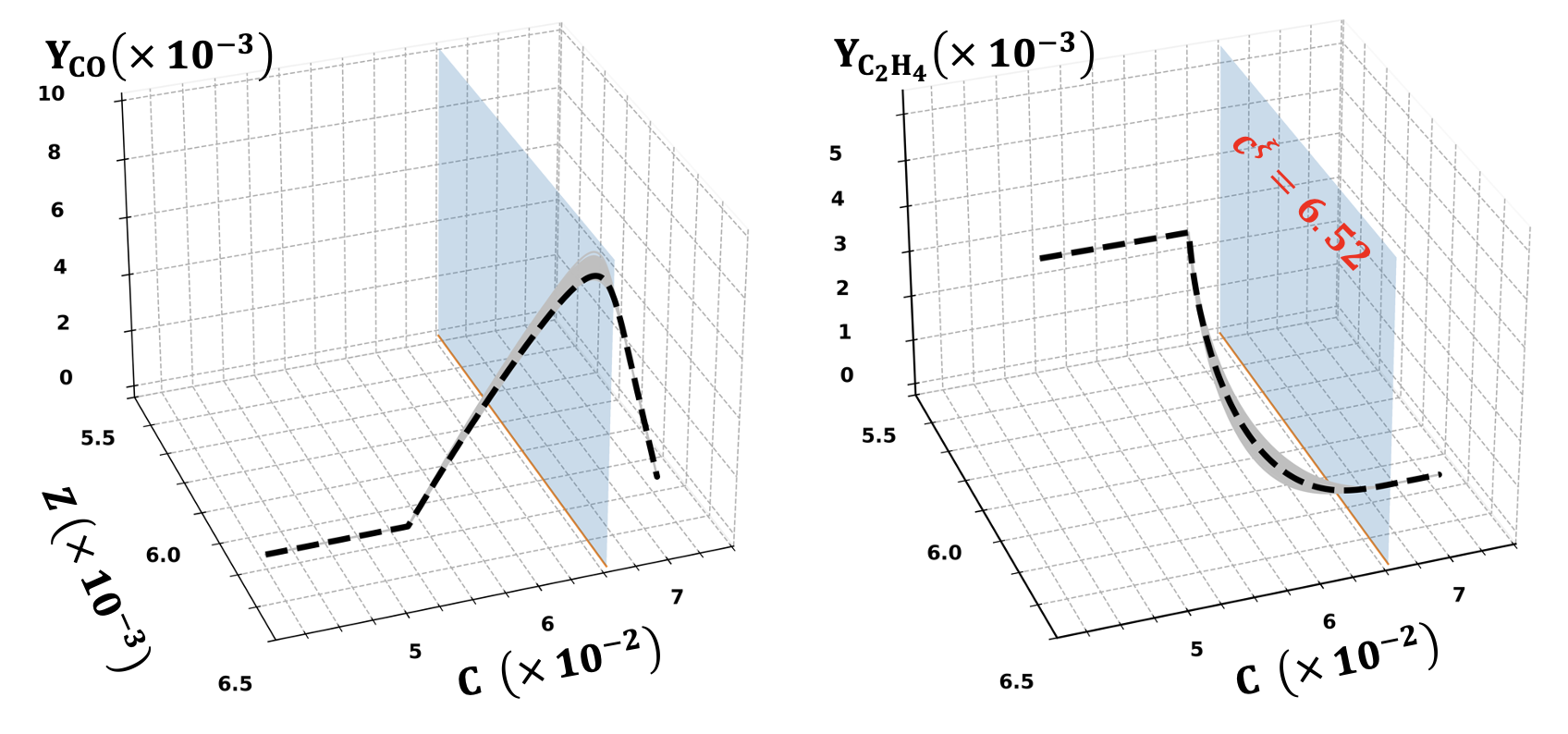}
    \caption{\footnotesize Ensemble of one-cell reaction trajectories in C, $Z$, $\mu$ space, and two different un-constrained coordinate definition. Gray curves span different sample of reaction kinetics; one typical trajectory is emphasized with a thick black dashed line. The blue plane marks  $C^{\xi}=0.0652$ constraint.}
\label{fig:onecell_traj}
\end{figure}

Following the procedure in $\S$\ref{sec:one_cell_mtm}, the ensemble of 500 different sampled reaction trajectories for the aforementioned cell is shown in Fig.~\ref{fig:onecell_traj}. To map the results to the simulation manifold $\boldsymbol{\xi}$, the reaction process was evaluated using the progress variable $C$ = Y$_{\mathrm{CO_2}}$ + Y$_{\mathrm{CO}}$, which is defined from the original cell state. The unconstrained coordinate (representing species that are not frozen during reaction evolution) is defined as the mass fraction of fuel and intermediate species. Intermediate species CO shows low-\(C\) induction with \(\mu \approx 0\), followed by a sharp increase as reactions accelerate and a subsequent decay of \(\mu\) as products accumulate while \(C\) continues to rise (a similar trend observed with other radicals, not shown here), but with a low standard deviation. For C$_2$H$_4$, all trajectories follow closely with different samples, indicating a lower sensitivity to the reaction rate parameters. The thermochemical states at the intersection point of each reaction trajectory with the plane \(C = c^{\xi}\) are used for next analysis.

\subsubsection{Local trajectory time\label{sec:simResults}} \addvspace{10pt}

\begin{figure}[!ht]
    \centering
    \includegraphics[width=0.95\linewidth]{ 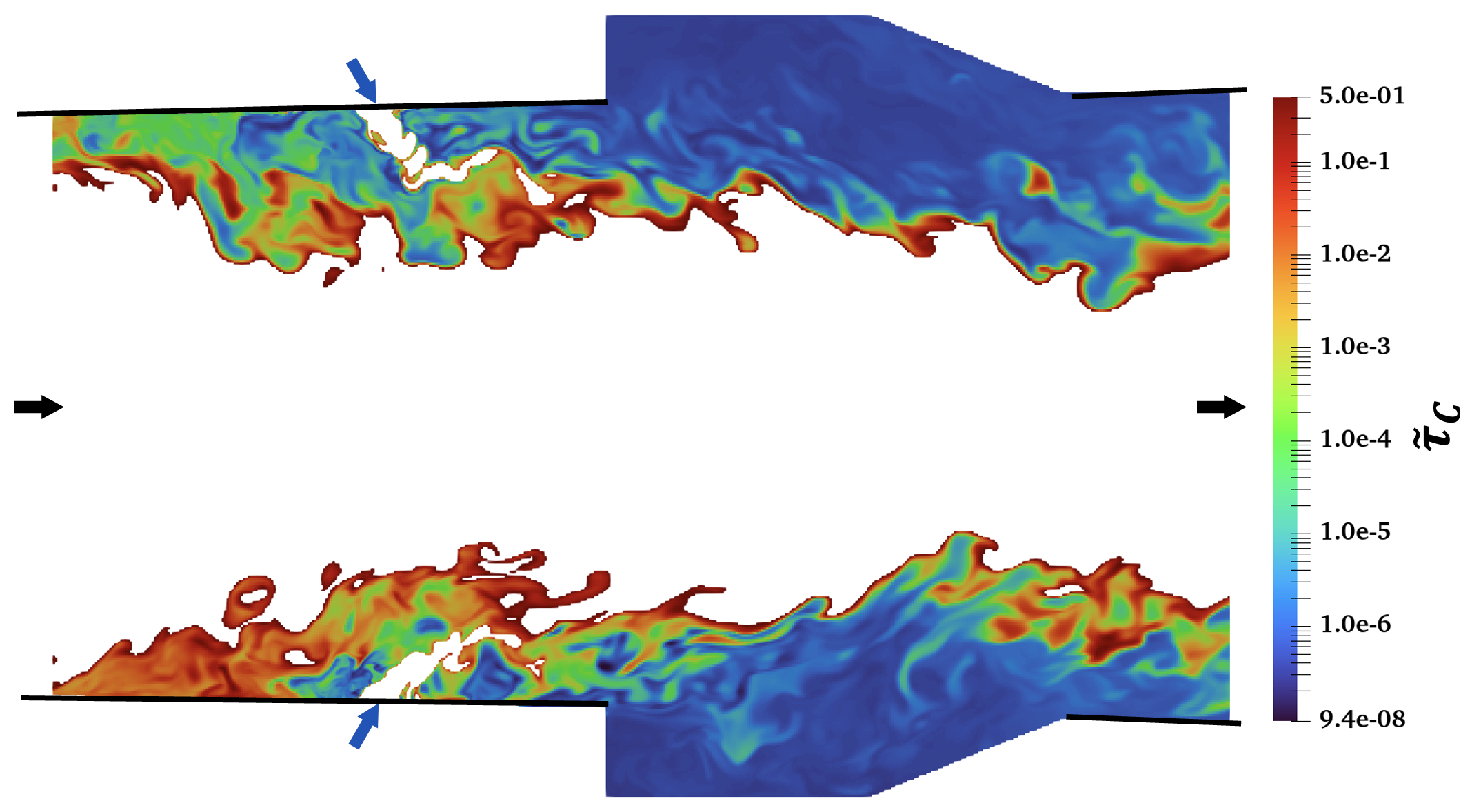}
    \caption{\footnotesize Instantaneous snapshot along the mid-section plane of the combustor for $\widetilde{\tau}_C$ (log-scale) contour computed from $\Psi$. Description of arrows in Fig.~\ref{fig:scrjet}. }
\label{fig:igdel}
\end{figure}

Following the definition in $\S$\ref{sec:simRes_MTM}, Fig.~\ref{fig:igdel} shows the mean time ($\widetilde{\tau_C}$) in the combustor mid-plane. The interactions between discrete injector mixing and recirculation lead to order-of-magnitude variations in timescales across the combustor. In Fig.~\ref{fig:igdel}, the regions of short ignition delay downstream and upstream of injection are governed by distinct, albeit related, behaviors. It should be noted that the ignition delay field is asymmetric despite an axisymmetric flow path due to the instantaneous nature of the analysis.

Within the active region downstream of injection, $\widetilde{\tau_C}$ reaches local minima on the order of 1e-06 s along the surface of the stoichiometric mixture fraction, i.e., the diffusion flame front~\cite{peters2000}. Because of the high convective velocities from the crossflow, the flame front maintains itself in the cavity where convective timescales are significantly increased. This allows for a similitude to mixing timescales, and temperatures and intermediate species build in the cavity and along the shear layer, reducing the overall ignition delay beyond the flame front.

The adverse pressure gradient from combustion backpressure encourages boundary layer separation, introducing a recirculation zone upstream of the injector as well~\cite{parker1995}. A less significant amount of fuel is entrained into this region. The relative leanness of the upstream mixture leads to $\widetilde{\tau_C}$ being up to a few orders of magnitude (1e-04 s) larger than that near the cavity. The subsonic recirculation width decreases with distance from the jet, and the $\widetilde{\tau_C}$ approaches infeasible burning conditions ($\sim$ 1e-03 s) as the fuel concentration approaches zero. Cells where the $\widetilde{\tau_C}$ reaches times that are unreasonable for the length of the geometry are overwritten to zero to represent an infeasible condition. For example, the high-temperature region behind the leading oblique shock in the core is visualized with zero $\widetilde{\tau_C}$. There is also a considerable region upstream of injection with relatively higher temperatures due to the viscous effects of the boundary layer rather than combustion.

\begin{figure}[!hb]
    \centering
    \includegraphics[width=0.99\linewidth]{ 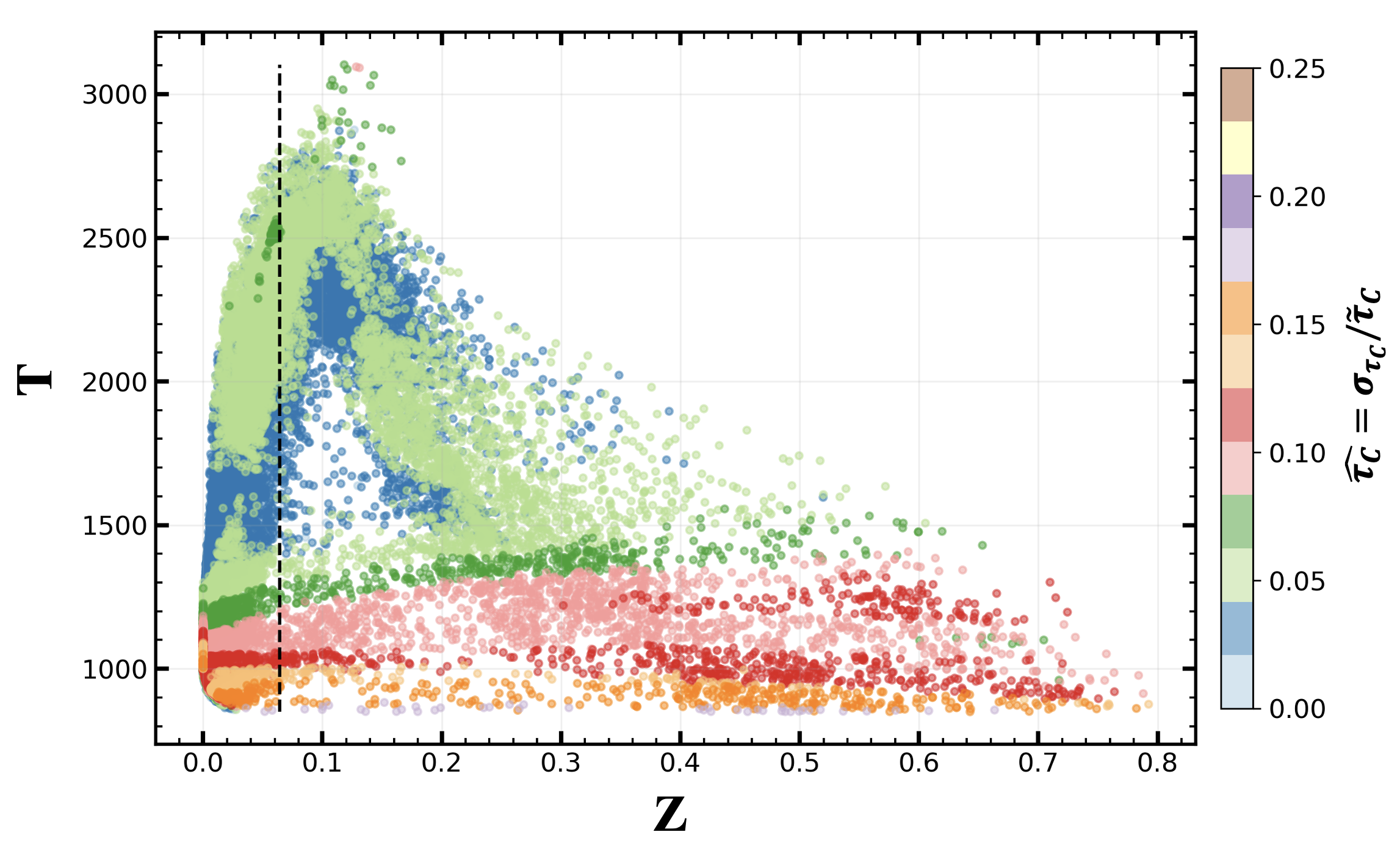}
    \caption{\footnotesize Scatter plot of T versus Z colored by $\widehat{\tau}_C$ and black-dash line indicating stoichiometric value of $Z_{st}$ = 0.0641. }
\label{fig:scatterC}
\end{figure}
Fig.~\ref{fig:scatterC} presents temperature plotted against mixture fraction, colored with variability in local timescale ($\widehat{\tau}_C$ = $\sigma_{\tau_C}$/$\widetilde{\tau_C}$). A consistent flamelet-type behavior is visible in the profile, while the color spread indicates higher variability near low T states. Across all entire Z range, for lower T, the variability is relatively higher (in the range of 17-10\%) compared to higher T states, where the variability is about 7-5\% (similar trends for DME in~\cite{su2021uncertainty}). Although $\widehat{\tau}_C$ changes gradually at a fixed $Z$, the UQ analysis shows that the time to reach a local reactive state exhibits pronounced dispersion at low initial temperatures (before initiating chain-branching)—conditions typical when ignition kernels interact with cold fuel streams.

\begin{figure}[!ht]
    \centering
    \includegraphics[width=0.99\linewidth]{ 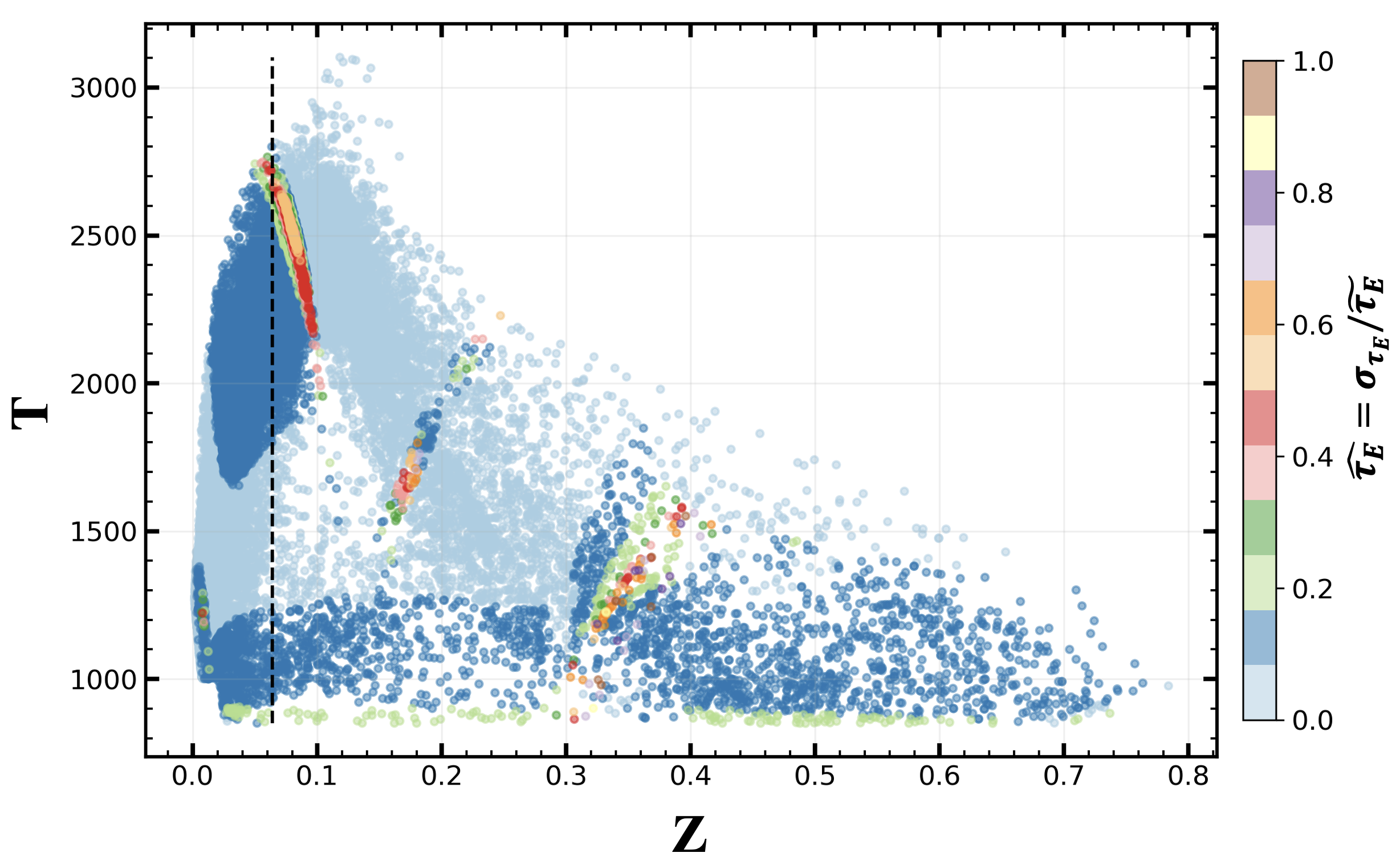}
    \caption{\footnotesize Scatter plot of T versus Z colored by $\widehat{\tau}_{E}$ and black-dash line indicating stoichiometric value of $Z_{st}$ = 0.0641. }
\label{fig:scatterUV}
\end{figure}

Figure~\ref{fig:scatterUV} shows the normalized variability in the equilibration time ${\tau_E}$ (see $\S$\ref{sec:simRes_MTM}). A similar trend is observed in the lower T and leaner Z region (well-mixed); however, a pronounced high-T ridge appears near the $Z_{st}$ dashed line (variability $\sim$ 55\%), with secondary bands at $Z\sim$ 0.2 and 0.3. The UQ results indicate equilibrium-shift reactions~\cite{wang1998comprehensive} specific to distinct (T,Z) regimes; modest perturbations to their rates produce order-of-magnitude spreads in local timescales. Although timescales are comparatively shorter at higher T, here thermal-pathways exhibit increased sensitivity to perturbations, which may critically influence or control other processes (e.g., phase change). 
%%% transport-limited processes

% -------------------------------------------------------------------- %
\section{Conclusion\label{sec:final}} \addvspace{10pt}

This work presents an applied uncertainty-mapping procedure that quantifies chemistry-induced variability on reduced manifolds by (i) reconstructing reduced-space states in the detailed mechanism’s full composition space, (ii) perturbing kinetic parameters using established covariance relations, and (iii) propagating each realization to a prescribed progress-variable level to obtain uncertainty estimates for ignition-relevant quantities.
The approach is first demonstrated in a subsonic, multi-tube combustor with interacting jet flames and strong recirculation. The resulting fields show order-of-magnitude variations in chemical timescale (${\tau_C}$) driven by jet–recirculation coupling and partially premixed stratification: near-stoichiometric regions along the flame front exhibit rapid kinetics ($\mathcal{O}(10^{-6})$s), while stratified zones near the tube exit and inter-jet recirculation produce substantially longer timescales (up to $\mathcal{O}(10^{-2})$s and spanning $\sim 0.2$–$10^{-4}$s). Extending the analysis to the time-to-equilibrium (${\tau_E}$) further localizes the largest variability ($\widehat{\tau}_E \approx 2\%$–$18\%$) to low-to-intermediate temperatures associated with induction and the onset of heat release, where competition among key HO$_x$ chain-propagation/branching pathways (R1–R3) governs relaxation and can influence ignition-kernel evolution in cooler, stratified streams.

The method is then applied to a high-speed flow case using a reduced mechanism (with FFCM-2 as the detailed reference). The resulting fields resolve the coupled effects of injector-driven mixing and upstream recirculation, showing rapid timescales near the stoichiometric diffusion flame and elevated variability ($\geq$10\%) in low-temperature regions. Equilibrium-timescale analysis further indicates that equilibrium-shift reactions can strongly influence regime-dependent timescales (with relative variation in $\widehat{\tau}_{E}$ exceeding 50\%). 
Embedding this applied UQ procedure within practical simulations offers a tractable means to identify where chemistry-driven uncertainty materially affects predicted behavior and where reduced mechanisms diverge from their detailed-mechanism reference. The framework produces spatially resolved, physically interpretable uncertainty maps for complex reacting flows and isolates operating regimes in which mechanism-reduction effects merit closer examination. Future work will derive detailed-mechanism UQ metrics and translate them into uncertainty maps for reduced-reaction mechanisms.

\acknowledgement{CRediT authorship contribution statement} \addvspace{10pt}
{\bf VS}: designed research, developed software, performed research, analyzed data, wrote—original draft, review and editing. 
{\bf SZ}: performed research, developed software, analyzed data, wrote—review and editing. 
{\bf RJ}: performed research, conducted simulations, analyzed data, wrote—review and editing. 
{\bf VR}: designed research, developed software, analyzed data, wrote—review and editing, project administration, funding acquisition.

\acknowledgement{Declaration of competing interest} \addvspace{10pt}
The authors declare that they have no known competing financial interests or personal relationships that could have appeared to influence the work reported in this paper.

\acknowledgement{Acknowledgments} \addvspace{10pt}
The authors acknowledge support from US office of Naval Research under Grant N00014-21-1-2475 with Dr. Eric Marineau as program manager.

% -------------------------------------------------------------------- %
% -------------------------------------------------------------------- %
% -------------------------------------------------------------------- %
\footnotesize
\baselineskip 9pt

% -------------------------------------------------------------------- %
% -------------------------------------------------------------------- %
% -------------------------------------------------------------------- %
\clearpage
\thispagestyle{empty}
\bibliographystyle{proci}
\bibliography{PCI_LaTeX}

% -------------------------------------------------------------------- %
% -------------------------------------------------------------------- %
% -------------------------------------------------------------------- %

\newpage

\small
\baselineskip 10pt

% -------------------------------------------------------------------- %
% -------------------------------------------------------------------- %
% -------------------------------------------------------------------- %

\end{document}